\documentclass[aps,prl,showpacs,twocolumn]{revtex4-1}

\usepackage{amsmath}
\usepackage{graphicx}
\usepackage{dcolumn}
\usepackage{bm}
\usepackage{color}
\usepackage{soul}

\begin{document}
\title{Fast and accurate modelling of nonlinear pulse propagation in graded-index multimode fibers}

\author{Matteo Conforti, Carlos Mas Arabi, Arnaud Mussot, and Alexandre Kudlinski}

\affiliation{Univ. Lille, CNRS, UMR 8523-PhLAM-Physique des Lasers Atomes et Mole\'cules, F-59000 Lille, France
\\Corresponding author: matteo.conforti@univ-lille1.fr}

\date{ \today}



\begin{abstract}
We develop a model for the description of nonlinear pulse propagation in multimode optical fibers with a parabolic refractive index profile. 
It consists in a 1+1D generalized nonlinear Schr\"odinger equation with a periodic nonlinear coefficient, which can be solved in an extremely fast and efficient way. The model is able to quantitatively reproduce recently observed phenomena like geometric parametric instability and broadband dispersive wave emission. We envisage that our equation will represent a valuable tool for the study of spatiotemporal nonlinear dynamics in the growing field of multimode fiber optics. 
\end{abstract}


\maketitle

Nonlinear pulse propagation in multimode fibers (MMFs) is focusing a tremendous research interest \cite{Picozzi2015} . Even if graded index (GRIN) MMFs have been invented long ago \cite{Gloge1973}, it is only very recently that the systematic and in depth study of the complex nonlinear spatiotemporal effects that can take place in these fibers has began \cite{Mafi2012}. The experimental observations of multimode solitons \cite{Renninger2013,Wright2015}, ultrabroadband dispersive waves \cite{Wright2015a,Wright2015b}, geometric parametric instability (GPI) \cite{Krupa2016,Eznaveh2017,Dupiol2017}, beam self-cleaning \cite{Wright2016,Krupa2017,Liu2016}, and novel forms of supercontinuum \cite{Krupa2016a,Lopez2016} are striking examples of the incredibly rich and complex scenario offered by nonlinear propagation in GRIN fibers.
The reason why these observations took so long to appear is that the study of spatiotemporal effect in MMFs is an intrinsically hard task from the experimental, theoretical and numerical point of view. 

The description of pulse propagation in MMFs must consider the three spatial and the temporal dimensions at the same time, because spatial and temporal effects cannot, in principle, be decoupled. Essentially two models are exploited for the mathematical description of propagation in MMFs: the 3+1D Generalized Nonlinear Schr\"odinger Equation (GNLSE) with a spatial potential, also named Gross-Pitaevskii Equation (GPE) in the context of Bose-Einsten condensates \cite{Krupa2016,Wright2015,Longhi2003,Renninger2013}, and the multi-mode GNLSE  (MM-GNLSE) \cite{Poletti2008}.  
In the GPE the transverse dimensions are accounted for in the propagation equation through the potential describing the refractive index profile of the fiber. In the MM-GNLSE the transverse dimensions are described indirectly, through the projection over the different fiber modes, which are coupled by the nonlinearity.
GPE is the most direct tool, but also the most computationally expensive: for example, the simulation of the propagation of multimode solitons over a few meters of fiber requires several days of computation \cite{Renninger2013}. The computational complexity can be partially reduced by considering exclusively radially symmetric modes \cite{Guizar2004}. The MM-GNLSE, consisting in $N$ coupled 1+1D GNLSE, permits to reduce the computational time only if a limited number of modes are excited (typically $N<10$), because the number of nonlinear coupling terms grows as $N^4$. When considering beams with a relative large size, taking into account only a few modes may lead to inaccurate results.

In this Letter we develop a model for the description of nonlinear pulse propagation in parabolic GRIN fibers. The model is derived from the 3+1D GPE with a spatial parabolic potential. It describes all the scenarios where a stable self-imaging pattern is generated, which is normally the case in parabolic MMFs. 
We obtain a 1+1D GNLSE with a periodic nonlinear coefficient, which can be solved in an extremely efficient way by standard split-step methods and requires very modest computational resources.
We show that our model accurately reproduces different phenomena peculiar to parabolic GRIN fibers, like geometric parametric instability and broadband dispersive wave emission. 

We start from the following form of Gross-Pitaevskii equation \cite{Krupa2016,Wright2015,Longhi2003,Renninger2013}
\begin{equation}\label{GPE}
i\partial_z E+\frac{1}{2\beta_0}\nabla_T^2 E + d(i\partial_t)E- \frac{\beta_0\Delta}{r_c^2}r^2E+\frac{\omega_0 n_2}{c}f_{NL}(E)=0,
\end{equation}
where $r^2=x^2+y^2$, $\nabla_T^2=\partial_x^2+\partial_y^2$ is the transverse Laplacian, $E$ is the field envelope expressed in $\sqrt{\mathrm{W}}/\mathrm{m}$, $\beta_0=\omega_0 n_0/c$, $n_0=n_{co}$ is the core refractive index (at the center of the fiber), $d(i\partial_t)=\sum_{n\geq 2}(i\partial_t)^n\beta_n/n!$ is the dispersion operator,  $\beta_n$ being the derivatives of the propagation constant at the carrier frequency $\omega_0$. The function
$f_{NL}(E)=(1+i\tau_s\partial_t)[(1-f_r)|E|^2E+f_r E\int h_r(t')|E(t-t')|^2dt']$ describes the Kerr and Raman nonlinear responses, and $\tau_s\approx 1/\omega_0$ is the self-steepening time.  

If we consider continuous-wave (cw) excitations, the propagation of a  beam in a parabolic GRIN fiber experiences self-imaging, due to the equal spacing of the propagation constant of the modes \cite{Mafi2012}. Self-imaging is a linear effect, but it is preserved also in presence of nonlinearity \cite{Karlsson1992}. Remarkably, it can be proved that there exist exact nonlinear and periodic propagation modes, which can be found as a self-similar transformation of any stationary nonlinear mode. The self-similar solution constructed from the fundamental fiber mode is linearly stable \cite{Longhi2004}. If the injected field is a Gaussian beam, in the weakly or moderate nonlinear regime, the field remains approximately Gaussian and the periodic variation of the beam parameters along the propagation coordinate can be calculated by exploiting a variational approach \cite{Karlsson1992}.
The amplitude of the solution takes the following form (the phase, which is not exploited in our derivation, is not reported)
\begin{align}
\label{Es}|E_s(x,y,z)|&=A_s\,|F_s(x,y,z)|=A_s\frac{a_0}{a(z)}\exp\left[-\frac{1}{2}\frac{r^2}{a^2(z)}\right],\\
a^2(z)&=a_0^2[\cos^2(\sqrt{g}z)+C\sin^2(\sqrt{g}z))],\label{az}
\end{align}
where $a_0$ is the beam spot size in $z=0$, $g=2\Delta/r_c^2$, $\Delta=(n^2_{co}-n^2_{cl})/2n^2_{co}$ is the relative refractive index difference, $A_s^2=2pn_0/(n_2\beta_0^2a_0^2)$, $C=(1-p)/(\beta_0^2a_0^4g)$, $p$ being a dimensionless number measuring the distance from beam collapse.

\begin{figure}[t]
\centering
\fbox{\includegraphics[width=\linewidth]{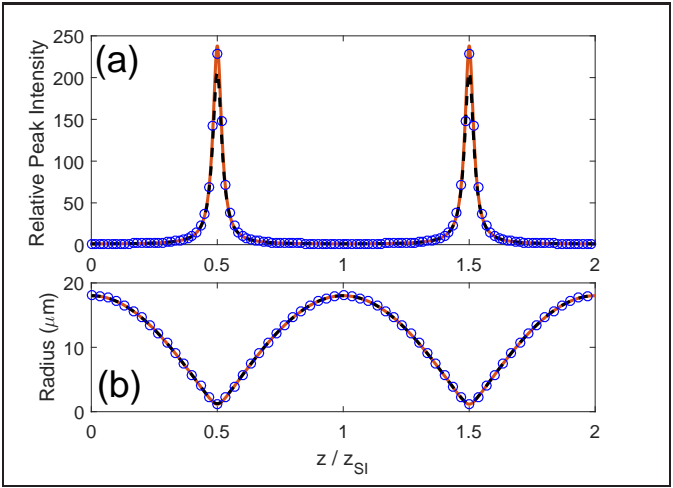}}
\caption{Gaussian solution of Eq. (\ref{GPE}). (a) Relative peak intensity $a_0^2/a^2(z)$ and (b) beam radius $a(z)$. Red curve, exact solution; black dashed curve, approximate linear solution; blue circles, numerical solution of GPE Eq. (\ref{GPE}) with $f_r=0$, $\tau_s=0$, $\partial_t=0$. Self-imaging period $z_{SI}=0.6155$ mm.}
\label{spatial}
\end{figure}

We make now two fundamental assumptions:  (i) the self-imaging pattern  remains stable during propagation \cite{Longhi2004}, and (ii) it is only slightly affected by nonlinearity. We thus write the solution of Eq. (\ref{GPE}) as
\begin{equation}\label{ansatz}
E(x,y,z,t)=A(z,t)\cdot F_s(x,y,z),
\end{equation}
where we have approximated the stationary self-imaging field with its linear shape [$p\approx 0$, $C\approx 1/(\beta_0^2a_0^4g)$ in Eq. (\ref{az})], and accounted for all the temporal and nonlinear effects in the envelope $A(z,t)$.
By inserting the Ansatz (\ref{ansatz}) into Eq. (\ref{GPE}), multiplying by $F_s^*$, and integrating over the transverse plane $x,y$, we get 
\begin{align}
 i\partial_z \psi + d(i\partial t)\psi+\gamma(z) f_{NL}(\psi)=0, \label{GNLS}\\ 
\gamma(z)=\frac{\omega_0n_2}{c A_{eff}(z)}=\frac{\omega_0n_2}{2\pi c\, a^2(z)},\label{gamma}
\end{align}
where $\psi(z,t)=A(z,t)\sqrt{\mathcal{S}}$ is the envelope normalized to the area $\mathcal{S}=\int\int|F_s(x,y,z)|^2dxdy=\pi a_0^2$, so that $|\psi|^2$ represents the optical power expressed in Watts \cite{Agrawal}, and $a^2(z)$ is given by Eq. (\ref{az}). Equation (\ref{GNLS}) constitutes a 1+1D GNLSE where the spatial effects are summarized by the periodic nonlinear coefficient Eq. (\ref{gamma}). Physically, the self-imaging pattern generates a $z$-varying effective area $A_{eff}(z)$, due to the periodic beam focusing, which thus couples the spatial evolution to the temporal envelope $\psi(z,t)$.


Equations (\ref{GNLS}-\ref{gamma}) are the main result of this Letter and in the following we show that they give a faithful reproduction of spatiotemporal effects in GRIN fibers.

For definiteness, in all the reported examples, we consider a standard GRIN MMF with a core radius  $r_c=26\,\mu \mathrm{m}$, core and cladding refractive index $n_{co}=1.470$, $n_{cl}=1.457$, giving a relative refractive index difference $\Delta=0.088$, and a nonlinear refractive index $n_2=3.2\cdot 10^{-20}\,\mathrm{m}^2/\mathrm{V}$ \cite{Krupa2016}.  We first show that the self-imaging pattern is not significanlty affected by the nonlinearity, in the range of powers typically exploited in the experiments, namely  a few hundreds of kilowatts  \cite{Wright2015,Wright2015a,Wright2015b,Krupa2016,Eznaveh2017,Dupiol2017}.
We consider the propagation of a cw Gaussian beam at 1064 nm centered on the input face of the fiber with a full width at half maximum (FWHM) intensity size of $30\,\mu \mathrm{m}$, and a peak power of $P_p=500$ kW. For the moment we neglect Raman and self-steepening effects ($f_r=0$, $\tau_s=0$). The relative peak intensity $|E(0,0,z)|^2/|E(0,0,0)|^2$ and the beam radius $a(z)$ are plotted in Fig. \ref{spatial}, calculated from analytical solution Eqs. (\ref{Es}-\ref{az}) and from numerical solution of stationary ($\partial_t=0$) GPE Eq. (\ref{GPE}). We first note that the Gaussian approximation \cite{Karlsson1992} (solid red curve) perfectly agrees with numerical solution of GPE (blue circles) over arbitrarily long distances \cite{Longhi2004} (a zoom over two self-imaging recurrences of period $z_{SI}=\pi r_c/\sqrt{2\Delta}=0.6155$ mm is shown in Fig. \ref{spatial}).
 Moreover, the approximate linear solution (dashed black curve) reproduces with great accuracy the exact one, proving the validity of our assumptions.
\begin{figure}[t]
\centering
\fbox{\includegraphics[width=\linewidth]{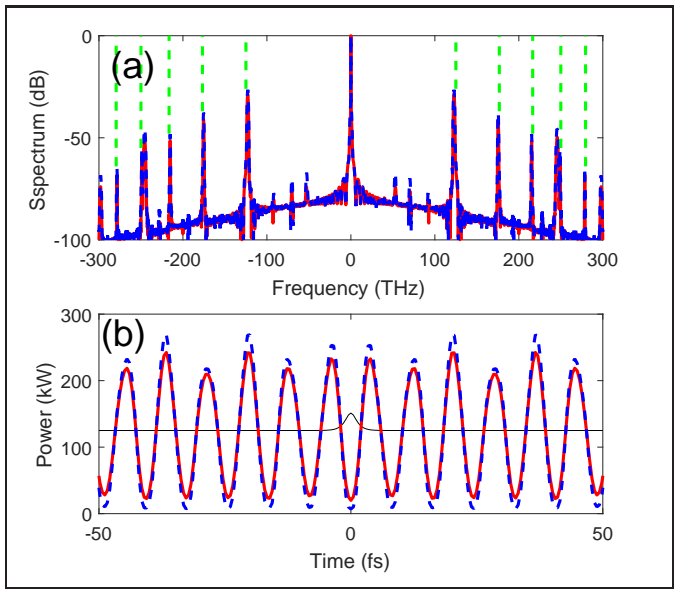}}
\caption{GPI arising after a propagation distance $z=5$cm of a Gaussian beam of 30 $\mu \mathrm{m}$ FWHM size and 125 kW peak-power. Output spectrum (a) and output temporal profile (b). Dashed green lines in (a) represent the GPI frequencies $f_m\approx \sqrt{m/(2\pi z_{SI}\beta_2)}$. Solid red curves, 1+1D GNLSE Eq. (\ref{GNLS}); Dashed blue curves GPE Eq. (\ref{GPE}). Thin black line in (b) is the initial temporal profile. Here $f_r=0$, $\tau_s=0$. }
\label{GPI}
\end{figure}

Equations (\ref{GNLS}-\ref{gamma}) perfectly reproduce complex spatiotemporal phenomena like GPI \cite{Longhi2003,Krupa2016}. 
In order to show this, we consider  the evolution of a cw Gaussian beam at 1064 nm with a FWHM beam size of $30\,\mu \mathrm{m}$ and peak power $P_p=125$ kW, in a fiber with dispersion $\beta_2=16.55\cdot 10^{-27}\,\mathrm{s}^2/\mathrm{m}$ and a pure Kerr nonlinearity, similarly to the conditions considered in \cite{Krupa2016}. GPI is a form of spatiotemporal modulation instability, which entails the generation of multiple sidebands at frequencies $f_m\approx \pm\sqrt{m/(2\pi z_{SI}\beta_2)}$, $m=1,2,\ldots$  \cite{Longhi2003}. In experiments GPI spontaneously emerges from random noise. Here, in order to facilitate a quantitative comparison and to get rid of any randomness in the initial condition, we seed GPI by adding a broadband coherent seed to the cw (a short hyperbolic secant pulse of duration 1 fs and one tenth of the cw amplitude). Figure \ref{GPI}(a) shows the spectrum after a propagation distance $z=5$ cm simulated by GPE (dashed blue curve) and 1+1D GNLSE (solid red curve). We first note that both methods reproduce the generation of GPI sidebands at the frequencies predicted by the theory (highlighted by dashed green lines). Most importantly, we point out the perfect agreement between the full GPE and the simple 1+1D GNLSE. The striking accuracy of our simplified model is confirmed by the comparison of the output temporal power profile reported in Fig. \ref{GPI}(b). As for the simulation time, GPE took roughly one hour, whereas 1+1D GNLSE only a few seconds (700 times faster). 
\begin{figure}[h]
\centering
\fbox{\includegraphics[width=\linewidth]{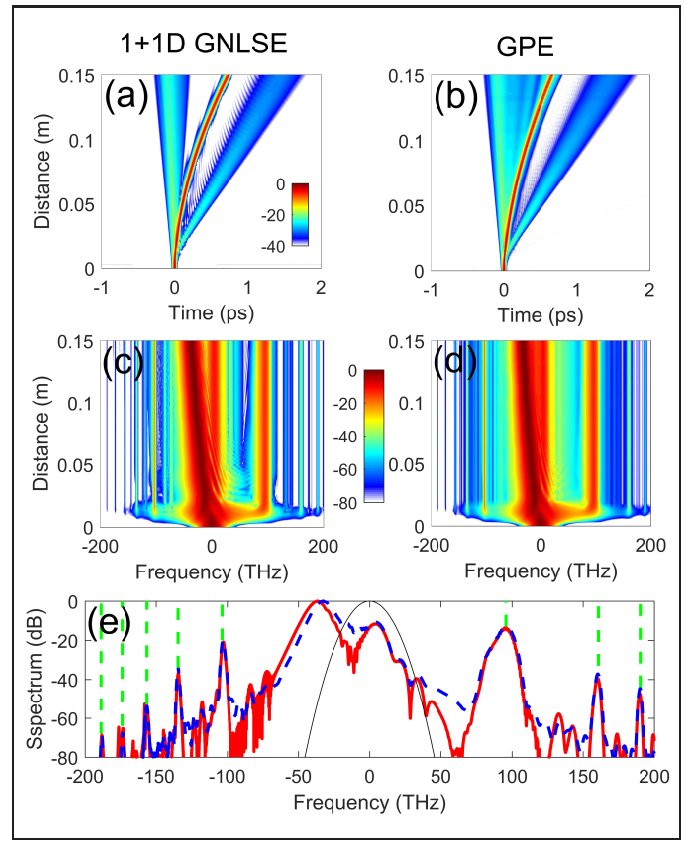}}
\caption{Emission of broadband dispersive radiation from a multimode soliton. (a-d) False color plots of the temporal (a,b) and spectral (b,d) evolution from 1+1D GNLSE (a,c) and GPE (b,d). (e) Input (thin black line) and output spectrum from 1+1D GNLSE (solid red curve) and GPE (dahsed blue curve). Vertical dashed green lines indicate the frequencies of the dispersive radiations from the Eq. (\ref{phasematching}). The inverse group velocity difference is estimated from (a) to be $\delta\beta_1=1.09$ ps/m. }
\label{soliton}
\end{figure}

Another spatiotemporal effect peculiar to GRIN fibers is the emission of ultrabroadband dispersive radiation from multimode solitons \cite{Wright2015a,Wright2015b}. We consider the propagation of a Gaussian pulse at 1550 nm with a FWHM size of 30 $\mu \mathrm{m}$, FWHM duration of 25 fs  and peak power $P_p=500$ kW. We consider second and third order dispersion $\beta_2=-22\cdot 10^{-27}\,\mathrm{s}^2/\mathrm{m}$, $\beta_3=1.32\cdot 10^{-40}\,\mathrm{s}^3/\mathrm{m}$,  Kerr and Raman nonlinearity ($f_r=0.18$, $\tau_s=0$). Figure (\ref{soliton})(a,b) shows the temporal evolution of the pulse power. A soliton is generated in the first millimeters of propagation, which sheds a dispersive wave packet and decelerates due to the Raman self-frequency shift. By looking at the spectrum [Fig. (\ref{soliton})(c-e)], it is evident that the dispersive wave is constituted by several spectral lines.
Indeed, the spatiotemporal pulsation of the soliton, which induces the nonlinear grating $\gamma(z)$, generates the polycromatic dispersive radiation according to the phase-matching relation \cite{Conforti2015,Conforti2016}
\begin{equation}\label{phasematching}
D(\omega)-\delta \beta_1\,\omega=\frac{2\pi}{z_{SI}}m+\gamma_{av}\frac{P_s}{2},\;\;m=0,\pm 1,\pm 2,\ldots
\end{equation}
where $\delta\beta_1$ arises from the deviation of the actual
group-velocity of the soliton from the natural one \cite{Conforti2013}, $D(\omega)=\beta_2\omega^2/2+\beta_2\omega^3/6$ is the dispersion operator in the frequency domain and $\gamma_{av}=n_2\omega_0\beta_0\sqrt{g}/(2\pi c)$ is the average nonlinear coefficient. This interpretation is well proved by Fig. (\ref{soliton})(e), where the spectral position of the dispersive waves at the end of the fiber is well predicted by Eq. (\ref{phasematching}). It is worth noting the impressive agreement between 1+1D GNLSE [Fig. (\ref{soliton})(a,c)] and GPE [Fig. (\ref{soliton})(b,d)], which permits the perfect superposition the output spectra shown in Fig. \ref{soliton}(e). We would like to stress again that the reduction of simulation time is dramatic: GPE tooks several hours, whereas 1+1D GNLSE less than one minute (500 times faster).

To conclude, we have derived a new model for the description of nonlinear propagation in parabolic GRIN fibers. We demonstrated that our model is as accurate as the full GPE for the simulation of complex spatiotemporal nonlinear dynamics such as geometric parametric instability and polychromatic dispersive wave emission from multimode solitons. In general, it is able to describe all the scenarios where a stable self-imaging pattern is generated inside the fiber. Given the drastic reduction of computational time of two orders of magnitude with respect to the full 3+1D equation, we expect that our model will become a workhorse for the description of nonlinear dynamics in MMFs. We also envisage that our equation will permit to discover new intriguing spatiotemporal phenomena, which are currently hidden by the computational burden required to solve GPE and MM-GNLSE.

\section*{Funding Information}
Agence Nationale de la Recherche (ANR) (ANR-11-EQPX-0017, ANR-11-LABX-0007, ANR-13-JS04-0004, ANR-14-ACHN-0014); CPER Photonics for Society P4S; IRCICA.

\end{document}